\documentclass[12pt]{iopart}
\usepackage{graphicx}
\begin{document}
\title{The motion of galaxy clusters in inhomogeneous cosmologies}
\author{David Garfinkle}
\address{Dept. of Physics, Oakland University,
Rochester, MI 48309, USA}
\ead{garfinkl@oakland.edu}
\begin{abstract}
Lema\^itre-Tolman-Bondi inhomogeneous spacetimes
can be used as a cosmological model to account for the type Ia 
supernova data.  However, such models also give rise to large velocities
of galaxy clusters with respect to the cosmic microwave background. These
velocities can be measured using the kinematic Sunyaev-Zeldovich effect.  This 
paper presents a calculation of galaxy cluster velocities as a function of
redshift for such a model.

\end{abstract}
\maketitle
\section{Introduction}

Observations of type Ia supernovae\cite{sn1,sn2}, 
along with the usual assumption that the universe is homogeneous, 
indicate that the expansion of the universe is accelerating.  This acceleration
necessitates the presence of an exotic form of matter called dark energy, which
can most easily be accounted for by a cosmological constant.  However, 
because the scale of the cosmological constant is so small 
($\sim {{10}^{-120}}$ in Planck units) it is natural to look for alternative 
explanations for the data.  One such alternative explanation is to account
for the supernova data using a comsmological model that is 
inhomogeneous.\cite{kolb1,kolb2}  
The simplest such models are the 
Lema\^itre-Tolman-Bondi(LTB) spacetimes.\cite{lemaitre,tolman,bondi} These
are spherically symmetric spacetimes with pressureless fluid (dust) as the 
matter. Models of this form can be constructed that account for the 
supernova data as well as the standard homogeneous ``concordance'' cosmology
does.\cite{flanagan,alnes,moffat,garcia0,garfinkle}  
Thus to test the difference between concordance cosmology and 
inhomogeneous cosmology, a different test is needed.  Such a test is 
provided by the kinematic Sunyaev-Zeldovich (kSZ)
effect.\cite{ksz}  This effect comes about
when cosmic microwave background (CMB) photons scatter off the gas in a galaxy 
cluster.  The spectrum of the scattered photons depends on the velocity
of the galaxy cluster with respect to the microwave background, and thus an
analysis of the CMB can be used to find this velocity. Thus, each LTB
model must be compared not only to the supernova data but also to the 
kSZ data, and only those models that match both sets of data can be 
considered viable.   

Analyses of the kSZ effect for particular LTB models have been 
performed\cite{caldwell,garcia} as well as a general analysis 
for slightly inhomogeneous
models.\cite{goodman} Unfortunately, each LTB model is different, and so
each must be seperately subjected to the kSZ test (as well as the supernova
test) to see if that model is viable. In this paper we will present a
general method for finding the velocity of galaxy clusters in LTB models.  
We will then apply this method to the models treated in\cite{garfinkle}.  
Section 2 presents the general method of calculating these velocities.  
Section 3 applies this method to the models of\cite{garfinkle}.  Conclusions
are presented in Section 4.

\section{calculating velocities}

The metric in the LTB spacetimes takes the form
\begin{equation}
d {s^2} = - d {t^2} + {\frac {{({r'})}^2} {1+f}} d {{\tilde r}^2} + 
{r^2} (d {\theta ^2} + {\sin ^2} \theta d {\phi ^2})
\end{equation}
Here the area radius $r$ is not one of the coordinates but is instead
a function of the coordinates $t$ and $\tilde r$.  The function $f$ is 
a function of $\tilde r$.  An overdot denotes derivative with respect to
$t$ while a prime denotes derivative with respect to $\tilde r$.  
The LTB spacetime has a prefered timelike vector 
${u^a}={{(\partial /\partial t)}^a}$ which is the four-velocity of the 
dust that makes up the matter in the spacetime.  However, an LTB spacetime
with a cosmic microwave background has another prefered vector, which we
will call $t^a$, the four-velocity of the CMB.  
Since the spacetime is
spherically symmetric, there is also a preferred unit spacelike vector
$r^a$ such that ${t^a}+{r^a}$ is an outgoing null vector and ${t^a}-{r^a}$
is an ingoing null vector. Expressed in terms of $t^a$ and $r^a$, the LTB
four-velocity $u^a$ takes the form
\begin{equation}
{u^a} = \gamma ({t^a} + v {r^a})
\end{equation}   
where $v$ is the velocity of the LTB observer with respect to the CMB and
where $\gamma =1/{\sqrt {1-{v^2}}}$. Now let $n_- ^a$  and $n_+ ^a$ be
respectively ingoing and outgoing CMB photons each with the same frequency
$\omega$ as measured in the CMB rest frame.  Then it follows that these
vectors take the form
\begin{eqnarray}
{n_- ^a} = \omega ({t^a} - {r^a})
\\
{n_+ ^a} = \omega ({t^a} + {r^a})
\end{eqnarray}
Now define the quantity $\alpha $ by 
\begin{equation}
\alpha = {\frac {{u_a}{n_+^a}} {{u_a}{n_- ^a}}} 
\end{equation}
Then it follows that 
\begin{equation}
v = {\frac {1-\alpha} {1+\alpha}}
\end{equation}
We restrict
attention to times when the CMB has decoupled from the matter, so that
$n_\pm ^a$ are null geodesics.  Thus, to find the velocity
of the galaxies with respect to the CMB, we need to find the behavior of
null geodesics in LTB spacetimes.

To find the behavior of null geodesics, 
we first recall some facts about the LTB spacetimes.
From the Einstein field equation, it follows that 
\begin{equation}
{{\dot r}^2} = f + {\frac F r}
\label{rdot}
\end{equation} 
where the function $F$ depends only on $\tilde r$.  The density is
given by 
\begin{equation}
\rho = {\frac {F'} {8 \pi {r^2} {r'}}}
\end{equation}
It is helpful to introduce the quantities $a,\, A$ and $B$ by
\begin{eqnarray}
r=a {\tilde r}
\\
f = A {{\tilde r}^2}
\\
F = B {{\tilde r}^3}
\end{eqnarray}
Then equation (\ref{rdot}) becomes 
\begin{equation}
{{\dot a}^2} = A + {\frac B a}
\label{adotsquared}
\end{equation} 
whose solution is 
\begin{equation}
t - {t_0} = {\int _0 ^a} {\frac {du} {\sqrt {A + {\frac B u}}}} 
\label{timediff}
\end{equation}
Here $t_0$ is a function of $\tilde r$ whose meaning is the time at which the
shell of dust with coordinate $\tilde r$ shrinks to zero radius.  To have
a genuine big bang singularity, we chose ${t_0}=0$.  We use the coordinate 
freedom
to change $\tilde r$ to any function of $\tilde r$ to set $B$ to the
constant $4/9$.  With this choice, the particular LTB model is then
determined by the choice of the function $A({\tilde r})$.  In 
particular, when $A=0$ the LTB solution becomes 
the standard dust FRW cosmology with $a={t^{2/3}}$.    

For any future directed null geodesic $k^a$ define the quantity
$\psi \equiv - a {k^a}{u_a}$.  Then it follows that 
${k^a} = - (\psi /a) {\ell ^a}$ where
\begin{equation}
{\ell ^a} = - {{\left ( {\frac \partial {\partial t}} \right ) }^a}  
+ \epsilon {\frac {\sqrt {1+f}} {r'}}  
{{\left ( {\frac \partial {\partial {\tilde r}}} \right ) }^a}
\label{elldef}
\end{equation}
where $\epsilon = \pm 1$ and the sign of $\epsilon$ depends on whether
$k^a$ is outgoing or ingoing.
Note from the definition of $\psi$ that 
\begin{equation}
\alpha = {\frac {\psi _+} {\psi _-}}
\end{equation}
where $\psi _+$ is the $\psi$ corresponding to $n_+ ^a$ and correspondingly
for $\psi _-$.  
It follows from the form of the LTB metric that  
\begin{equation}
{k^a}{k^b}{\nabla _a}{u_b} = {\frac {{\psi ^2} {{{\dot r}'}}} {{a^2} {r'}}}
\end{equation}
Then using the geodesic equation we find
\begin{equation}
{k^a}{\nabla_a} \psi = - {\frac {\psi ^2} {{a^2}{r'}}} ( - {r'} {\dot a}
+ a {{\dot r}'} + \epsilon {\sqrt {1+f}} {a'} )  
\end{equation}
from which it follows that 
\begin{equation}
{\ell ^a} {\nabla _a} \ln \psi = {\frac 1 {a {r'}}} ( {\tilde r} 
[ a {{\dot a}'} - {\dot a} {a'} ] + \epsilon {\sqrt {1+f}} {a '})
\label{main}
\end{equation}

we will use equation (\ref{main}) to find the velocity $v$ of the 
dust relative to the CMB as follows: note that since we have chosen
${t_0}=0$ it follows that at early times the spacetimes becomes FRW.  
Note also that in FRW spacetimes ${a'}=0$ and therefore $\psi$ is constant.
For CMB photons with the average energy this constant, which we will
call $\psi _0$ is the same for outgoing and ingoing photons.  Integrating
equation (\ref{main}) back from the point at which we want to calculate
$v$ to the big bang yields $\ln (\psi/{\psi _0})$ and since we do this
for both outgoing and ingoing geodesics, we obtain both 
${\psi _+}/{\psi _0}$ and ${\psi _-}/{\psi _0}$ which in turn allows us
to calculate ${\psi _+}/{\psi _ -}$ and thus the velocity $v$.      

We will numerically integrate equation (\ref{main}).  However, in order
to do that we will need to be able to evaluate all the quantities on
the right hand side of that equation at all points in the integration.  
As we will show, all that is needed is to know $r$ and $a$ and the rest of the
quantities follow.  However, $r$ and $a$ also change as the integration
procedes, so we need to know how they change.  
From equation(\ref{elldef}) we find
\begin{eqnarray}
{\ell ^a}{\nabla _a} r = \epsilon {\sqrt {1+f}} - {\dot r} 
\label{evolver}
\\
{\ell ^a} {\nabla _a} a = \epsilon 
{\sqrt {1+f}} {\frac {a'} {r'}} - {\dot a} 
\label{evolvea}
\end{eqnarray}
Note that an ingoing geodesic that goes through $r=0$ becomes an outgoing
geodesic, where this change is implemented as a change in the sign of 
$\epsilon$.
We must evolve the system of equations (\ref{main}-\ref{evolvea}).  To
do this, we must be able to evaluate all quantities on the right hand
sides of these equations from just $r$ and $a$, that is, we must be
able to find ${\tilde r}, \, f, \, {\dot a}, {\dot r}, {a'}, \, {r'}, \,
{{\dot a}'}$ and ${\dot r}'$.  Given $r$ and $a$ we obtain
${\tilde r}=r/a$ and thus we find all specified functions of $\tilde r$
such as $A$ and $f$.  From equation (\ref{adotsquared}) we obtain
\begin{equation}
{\dot a} = {\sqrt {A + {\frac B a}}}
\label{adot}
\end{equation}
which then yields $\dot r$ since ${\dot r} = {\dot a}{\tilde r}$.
Differentiating equation
(\ref{timediff}) with respect to $\tilde r$ and solving for $a'$ yields
\begin{equation}
{a'} = {\dot a} I {A'}
\end{equation} 
where 
\begin{equation}
I = {\frac B {2{A^{5/2}}}} \left [ s \left ( 3 + {\frac {Aa} B} \right )  - 
{\frac 3 2} \ln \left ( {\frac {1+s}
{1-s}} \right ) \right ] 
\end{equation}
with $s \equiv {\sqrt A}/{\dot a}$.  
This then yields $r '$ through ${r '} = a + {\tilde r}{a'}$.
Finally, differentiating equation (\ref{adot}) with respect to $\tilde r$
yields
\begin{equation}
{{\dot a}'} = {\frac {A'} 2} \left ( {\frac 1 {\dot a}} - 
{\frac {B I} {a^2}} \right ) 
\end{equation}
which in turn yields ${\dot r}'$ through 
${{\dot r}'} = {\dot a} + {\tilde r}{{\dot a}'}$.   

What remains is then simply to find the initial values of $r$ and $a$ for 
each point in our past light cone, along with the coresponding values of 
the redshift $z$.  As shown in \cite{garfinkle} this can be done by integrating
equations (\ref{evolver}-\ref{evolvea}) from our current position, along with
the evolution equation for the redshift
\begin{equation}
{\ell ^a}{\nabla _a} z = {\frac {{\dot r}'} {r'}} (1+z)
\end{equation} 
For this integration, the initial value of $r$ is zero, while the initial value
of $a$ is given by 
\begin{equation}
a = {\textstyle {\frac 4 9}} ( {\Omega _M ^{-1}} - 1)
\end{equation}
where $\Omega _M$ is the ratio of the density in matter to the 
critical density.  

\section{results}

In the models of \cite{garfinkle} a choice of $\Omega _M$ is made, and 
the function $A$ takes the form
\begin{equation}
A = {\frac 1 {1+{{(c {\tilde r})}^2}}}
\end{equation}
where the constant $c$ is chosen for best fit with the supernova data.  
In particular, the two models considered are (i) ${\Omega _M} = 0.3$, 
which gives rise to $c=8.5$, and (ii) ${\Omega _M} = 0.2$, which gives
rise to $c=5.1$.  Each of these models fits the supernova data about as
well as the standard concordance cosmology.  Because $A \to 0$ at large
$\tilde r$, these models are asymptotically $\Omega =1$ FRW cosmologies 
with an underdensity in a region near the center.  Note that the large value
of $c$ means that in these models we live in an underdensity of fairly
small size.  

Using the method of the previous section, we calculate the velocity of the
dust with respect to the CMB for each point on our past light cone.  The
results are shown in figures (1) and (2).  Here figure (1) is for the 
${\Omega _M} = 0.3 $ model and figure (2) is for the ${\Omega _M} = 0.2 $
model. Also plotted on the figures are the results of observations of 
the kSZ effect for galaxy clusters.\cite{szo1,szo2,szo3} 
\begin{figure}
\includegraphics{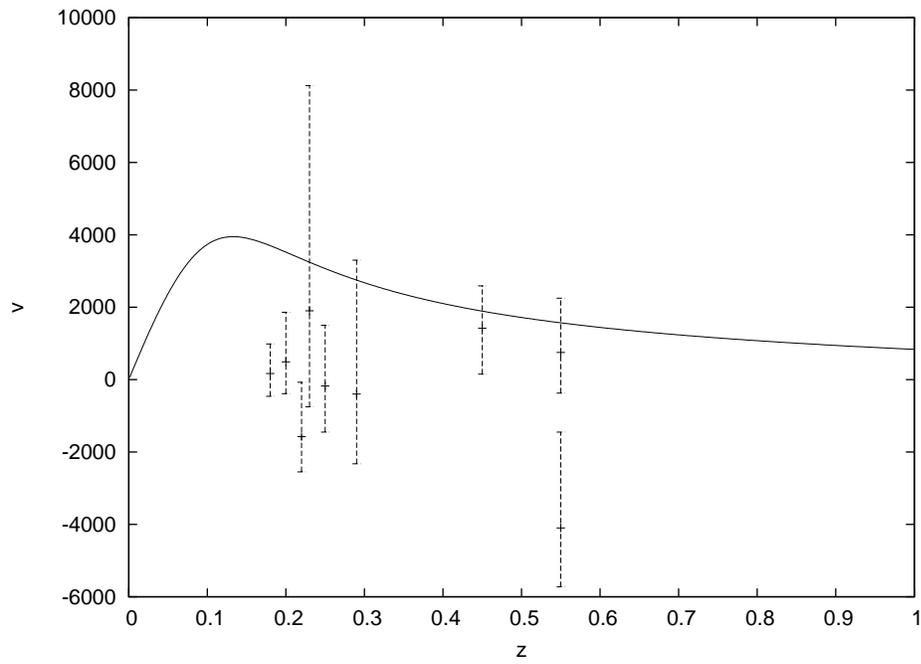}
\caption{\label{fig1}Plot of velocity (in units of km/s) 
{\it vs} redshift for the
${\Omega _M}=0.3$ LTB model }
\end{figure}
\begin{figure}
\includegraphics{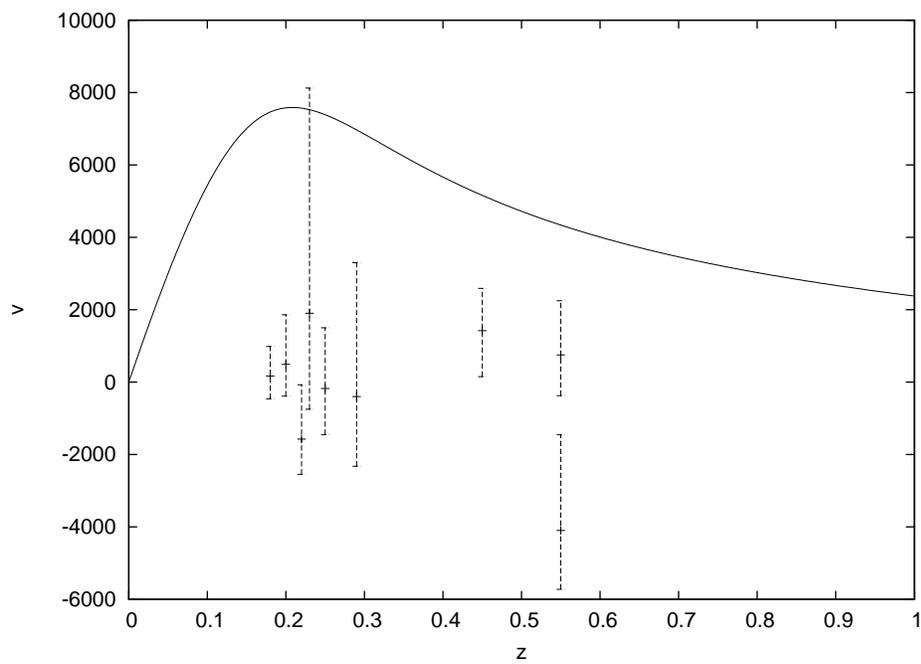}
\caption{\label{fig2}Plot of velocity (in units of km/s)
{\it vs} redshift for the
${\Omega _M}=0.2$ LTB model }
\end{figure}

The observations clearly rule out the ${\Omega _M}=0.2$ model.  In addition,
the ${\Omega _M}=0.3$ model does not provide a good fit to the observations,
so this model may also be ruled out.

\section{conclusions}

Any successful LTB model must pass both the supernova test and the kSZ test.
Together these tests provide a powerful constraint that rules out the 
models of \cite{garfinkle}.  However, the models of \cite{garfinkle} were 
chosen on the basis of simplicity and fitting the supernova test; so it is
not surprising that one additional test suffices to rule them out.  It is
therefore possible that reasonable LTB models could be made that
have fair agreement with 
both the supernova data and the kSZ data.  However, as kSZ measurements
improve in accuracy, this test will become more stringent and might eventually
rule out all LTB models.  

\ack
I would like to thank Nick Kaiser and Juan Garcia-Bellido for helpful 
discussions.  This work was supported by NSF grants PHY-0456655 and
PHY-0855532     
to Oakland University.

\end{document}